\documentclass[review]{elsarticle}

\usepackage{lineno,hyperref}
\usepackage{adjustbox}
\usepackage{graphicx,multirow}
\usepackage{listings}
\usepackage{xcolor} 
\usepackage{lmodern}  
\usepackage{amsmath}  
\usepackage{algorithm}  
\usepackage{subcaption}

\journal{Simulation Modelling Practice and Theory}

\makeatletter
\def\@author#1{\g@addto@macro\elsauthors{\normalsize%
		\def\baselinestretch{1}%
		\upshape\authorsep#1\unskip\textsuperscript{%
			\ifx\@fnmark\@empty\else\unskip\sep\@fnmark\let\sep=,\fi
			\ifx\@corref\@empty\else\unskip\sep\@corref\let\sep=,\fi
		}%
		\def\authorsep{\unskip,\space}%
		\global\let\@fnmark\@empty
		\global\let\@corref\@empty  
		\global\let\sep\@empty}%
	\@eadauthor={#1}
}
\makeatother

\lstdefinestyle{mystyle}{
	basicstyle=\scriptsize\ttfamily,
	breakatwhitespace=false,         
	breaklines=true,       
	captionpos=b,                    
	columns=fullflexible,
	frame=single,
	numbers=left,                    
	numbersep=5pt,         
	numberstyle=\tiny,          
	showspaces=false,                
	showstringspaces=false,
	showtabs=false,                  
	tabsize=2
}

\usepackage{tikz}

\DeclareFixedFont{\ttb}{T1}{txtt}{bx}{n}{12} 
\DeclareFixedFont{\ttm}{T1}{txtt}{m}{n}{12}  

\usepackage{color}
\definecolor{deepblue}{rgb}{0,0,0.5}
\definecolor{deepred}{rgb}{0.6,0,0}
\definecolor{deepgreen}{rgb}{0,0.5,0}

\usepackage{listings}

\newcommand\pythonstyle{\lstset{
		language=Python,
		basicstyle=\ttm,
		otherkeywords={self},             
		keywordstyle=\ttb\color{deepblue},
		emph={MyClass,__init__},          
		emphstyle=\ttb\color{deepred},    
		stringstyle=\color{deepgreen},
		frame=tb,                         
		showstringspaces=false            %
}}

\lstnewenvironment{python}[1][]
{
	\pythonstyle
	\lstset{#1}
}
{}

\newcommand\pythoninline[1]{{\pythonstyle\lstinline!#1!}}

\begin{document}

\begin{frontmatter}

\title{YAFS: A simulator for IoT scenarios in fog computing}

\author[mymainaddress]{Isaac Lera\corref{mycorrespondingauthor}}
\cortext[mycorrespondingauthor]{Corresponding author}
\ead{isaac.lera@uib.es}

\author[mymainaddress]{Carlos Guerrero}

\author[mymainaddress]{Carlos Juiz}

\address[mymainaddress]{Department of Mathematics and Computer Science, University Of Balearic Islands, Palma, 07122, Spain}

\begin{abstract}
We propose a fog computing simulator  for  analysing the design and deployment of applications through customized and dynamical strategies. We model the relationships among deployed applications, network connections and infrastructure characteristics through complex network theory, enabling the integration of topological measures in dynamic and customizable strategies such as the placement of application modules, workload location, and path routing and scheduling of services.  We present a comparative analysis of the efficiency and the convergence of results of our simulator with the most referenced entity, iFogSim. To highlight YAFS functionalities, we model three scenarios that, to the best of our knowledge, cannot be implemented with current fog simulators: dynamic allocation of new application modules, dynamic failures of network nodes and user mobility along the topology. 

\end{abstract}


\end{frontmatter}

\section{Introduction}

Cisco coined the term ``fog computing'' as an extension of cloud computing, placing computer services closer to the users~\cite{bonomi2012fog,DefOF,mell2011nist}.
Approximately speaking, some network devices, called fog nodes, perform computational tasks or data storage functions in the same way as cloud entities. This novel application placement has some advantages, such as the reduction of latency time,  a lower network bandwidth utilisation, a reduction in the cloud costs, and an increase in the reliability and fault  tolerance through the geographical distribution of devices.

 Another related concept similar to fog computing is edge computing~\cite{edgecomputing,EdgeEmergence,DO2018}. The small difference between them lies in the localization of  fog nodes. In edge computing, the nodes are at the edge of the network, near  the users.  In the context of the Internet of Things (IoT), the role of fog computing is to leverage functionalities such as on-demand scalability, real-time interaction, better security and privacy management, battery power savings, streamlining of communications, and rapid service delivery, among others~\cite{edgeiot,Buyya:2016:ITP:3050877,EdgeEmergence}.  
Fig.~\ref{fig1} shows the topology of a network with typical en-routing entities in comparison to network nodes with computational and storage capabilities in the fog computing model. 	The workload requests are generated from the  \emph{endpoints / things} layer and are routed to services deployed in upper-layer devices. These services can be allocated  in several intermediate nodes or in the cloud.


The placement problem of software resources for fog computing or edge computing is an NP problem that consists of the selection of the optimal network entity to deploy a user application. There are some constraints and optimization factors that influence this problem,  such as the user location, hardware and software features of the network entities, link characteristics (e.g., propagation, utilisation, and bandwidth), user requirements, application decomposition (e.g., containers, microservices, and serverless functions), QoS, energy, and cost, among others. 
These factors affect the dynamical evolution of the user movement, link failures, network congestion, and application popularity, among other aspects.  One way to evaluate placement solutions is through simulation. Simulators are enabling tools for modelling, analysing and evaluating the diversity of policies and configurations. 

In this paper, we present a  discrete event simulator focused on, but not restricted to, fog environments called YAFS (Yet Another Fog Simulator). YAFS is designed to analyse the design of applications and incorporates strategies for placement, scheduling and routing.  We compare the YAFS characteristics with that of the iFogSim~\cite{ifogsim} simulator as a reference.  iFogSim is the most widely used fog simulator, an extension of the well-known CloudSim~\cite{cloudsim} simulator. YAFS includes more functionalities than current simulators for modelling  IoT scenarios. We highlight the following points:

\begin{figure}
	\centering
	\includegraphics[width=.8\linewidth]{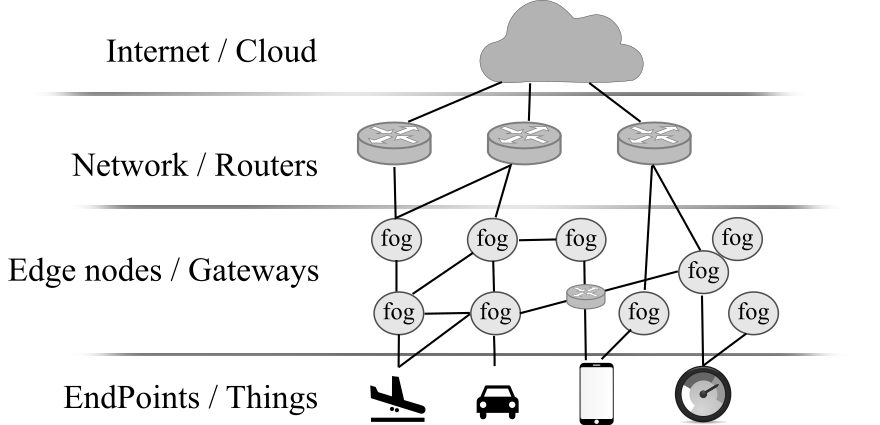} 
	\caption{Relationships in a fog computing model.}
	\label{fig1}
\end{figure}%

I) \emph{Network}: the network is modelled through complex networks enabling the incorporation of topological features in the placement solution.  A complex network is a graph with non-trivial topological features~\cite{Newman2003}. Complex network theory links  studies in several research fields through mathematical analysis of graph features~\cite{complex-networks-strogatz} such as clustering, assortativity, communities, communicability, flows, isomorphism, and similarity. In the literature, complex networks are used to model relationships of datacentres~\cite{filiposka2014complex}, fog colonies~\cite{Lera18Iot}, load balancing mechanisms~\cite{zhang2010load}, and so on. We can take advantage of extensively tested implementations to use them inside of custom simulator policies.  In our network modelling, fog devices and links can be created or removed throughout the simulation execution according to custom temporal distribution and can be  freely tagged to enable efficient and flexible definition of scenarios. The simulator controls the transmission of application messages and reports the communication failures. It provides raw data to compute the total number of unsatisfied requests, the best placements in failure cases, and the design of robust networks, among other outputs.  The simulator also supports CAIDA~\cite{CAIDA} and BRITE~\cite{BRITE} topologies.

II) \emph{Workload sources}: each workload source represents the connection of a user or an IoT sensor or actuator that demands a service. Each source is associated with a network entity and generates requests according  to a custom distribution. The workload sources can be created, changed or removed dynamically enabling the modelling of the user movements in an ecosystem. 

III) \emph{Placement, scheduling and routing custom algorithms}: these algorithms are defined by the user. The placement algorithm is invoked in the initialisation and runs along with the execution according to a custom distribution.  The routing algorithm chooses the path that connects the transmitter and the receptor, and the scheduling algorithm chooses the application that runs the task.  The scheduling and routing algorithms are defined in the same manner since the path selection depends on which application is chosen.  The existence of the scheduling algorithm allows choosing between different modules in case of scaling policies. By default, the simulator includes both implementations: a static placement and the selection of the minimum path between two entities where the application is deployed.

IV) \emph{Custom processes}: custom functions can be invoked at runtime to provide flexible implementations of real events such as the movement of the workload sources, generation of network failures, and specific data collection using third-applications such as Grafana~\cite{grafana} (an open platform for analytics and monitoring of computer infrastructures and services).

V) \emph{Post-simulation data analysis}: YAFS performs automated CSV-based logging of two types of events: workload generation and computation, and link transmissions. The results are analysed post-simulation,  which generates less overhead, avoids repeating the simulation to re-analyse other indicators, and enables the shareability of raw results. YAFS  includes functions to obtain metrics  such as network utilisation, response time, network delay, and waiting time, and other data analysis can be implemented by the user, such as the number of requests above QoS requirements or the peak of transmissions in a link.

VI) \emph{JSON-based scenario definition}: YAFS supports the importation of the scenario definition from JSON-format files. It enables the use of third-party  tools that generate scenarios in a common JSON format; in addition,  non-expert developers can use basic functionalities of the simulator.


YAFS is developed in Python following the style guide PEP8~\cite{p8p}. It is available under MIT licence in a code repository\footnote{https://github.com/acsicuib/YAFS} with detailed documentation, a tutorial and several  examples.


The contribution of this paper is the design of a highly customisable and adaptable simulator and the design of JSON-based files for analysing mobile IoT scenarios under the fog and edge computing paradigms.

This paper is organised as follows: Section 2 describes the state the art of the fog and cloud computing simulators. Section 3 includes the justification of the use of complex networks to model the infrastructure network. Section 4 describes the design and some details of the implementation of each component of our proposal. Section 5 includes three cases studies (involving allocation of modules, failure behaviour of devices, and movement of users) and presents a comparative study of results with iFogSim in terms of performance and  convergence.

\section{Related Work}

This approach is focused in the design of fog computing simulators. There are different simulators for several types of distributed environments such as cloud, grid, and fog edge. We know of four specific simulators regarding the topic of fog computing:  FogTorch~\cite{FogTorch}, EmuFog~\cite{emufog}, EdgeCloudSim~\cite{EdgeCloudSim} and iFogSim~\cite{ifogsim}. We first analyse some common features of these simulators and then provide  more specific features of all of them individually.

We classify the simulators considering  the following criteria, which, from our point of view, are essential for realistic modelling of fog scenarios.
I) The first criterion involves the structure of the topology. The topology allows us to represent the infrastructure of the network. EdgeCloudSim and IFogSim  use a hierarchical structure. In contrast,  FogTorch, EmuFog and YAFS  use a graph structure.  In addition, YAFS supports the definition of subgraphs within a topology. Subgraphs can be used to represent isolated regions or fog colonies~\cite{Lera18Iot}.  The generation of topologies is a complex and hard task due to the number of elements and connections. There are definitions of topologies such as BRITE and CAIDA topologies. These formats are supported by EmuFog and YAFS but not by the other simulators.
II) The second criterion is related to the coding of the scenario. Existing fog simulators include an API where the characteristics of the scenario are defined; YAFS also supports the definition of the scenario through JSON-based files. The topology can also be defined using this syntax. 
III) The third criterion involves the characteristics of the results. If complex scenarios are designed with customizable policies, it is necessary to record all the events of the simulator in files. Thus, users can perform complex analysis of these records to find specific indicators. EdgeCloudSim and YAFS record these data to allow post-simulation analysis.
IV) The fourth criterion involves the capability to perform changes in the fog scenario during the simulation. The modelling of realistic scenarios must include changes in the different strategies. 
EdgeCloudSim supports changes representing the movement of users in the infrastructure. Only YAFS supports dynamic  scenarios in the next strategies: placement, path routing, service orchestration and workload or user movement. 
V) Finally, the last criterion is the programming language. All other simulators are implemented in Java, but YAFS uses Python. 

We summarize these five comparative criteria in Table~\ref{tablarw}. The nomenclature used in the \emph{Policy} column is related to the specific policy used: \emph{A} means allocation or placement; \emph{R} - path routing,  \emph{O} - service orchestration, and \emph{W} - workload or user movement. \emph{Dynamic Policies} indicates those simulators that can change their strategies dynamically during execution.  
	
	\begin{table}[] 
		\centering
		\scalebox{0.65}{     
			\begin{tabular}{r|p{1.5cm} p{1.5cm} p{2cm} p{2cm} p{1.7cm} c p{1.5cm}}
				Simulator &    Coding &  Topology Structure  & Topology  definition & Result traceability & Policy  & Dynamic Policies  & Language    \\\hline\hline
				FogTorch          & API & graph & API & N& [A] & -    &  Java   \\ \hline
				EmuFog            &API & graph & API, Brite, CAIDA, \ldots & N& [A] &  -  &  Java \\\hline
				EdgeCloudSim  &API & tree & API & Y &  [A,W] & W  & Java \\\hline
				iFogSim            &API & tree & API & N& [A]  & -  & Java \\\hline
				\textbf{YAFS}                & API, JSON & graph & API, JSON, Brite, CAIDA,\ldots & Y & [A,R,O,W]  & all & Python \\\hline
			\end{tabular}
		}
		\caption{Comparative table of fog simulators.}
		\label{tablarw}
	\end{table}

After a global analysis of the simulators, we also present a more specific description for each of them independently.

FogTorch~\cite{FogTorch} uses Monte Carlo simulations to determine the best allocation for an application through QoS indicators such as latency, bandwidth, cost, and response time. This simulator addresses the application allocation problem. Our approach simulates the whole ecosystem  only where the allocation is one of the available inputs of the simulation. In other words, FogTorch optimizes the deployment of applications under QoS restrictions, and YAFS integrates this optimized allocation values to obtain simulated metrics.  Brogi et al. defined an application as a set of triplets of software components and interactions among components with a QoS profile. They used Monte Carlo simulations to compute the eligible deployments of software components. They also presented a fire alarm IoT application as a case study with three components: a fire manager (an actuator to extinguish the fire), a database system, and a machine learning engine. The IoT infrastructure was based on three fog nodes, two cloud entities and nine network links among them. 

EmuFog~\cite{emufog} is a set of scripts to transform a set of initial configurations (network topology and placement criteria) into the input of the MaxiNet~\cite{wette2014} simulator. It uses a graph representation to define the network topology. The authors implemented some functionalities to simplify the process of selection of fog nodes in regards to the topological features of the graph. Our simulator also implements this process. We delegate this type of computational processes in a complex network library to obtain topological features that the user can integrate into the topology. EmuFog application representation comes from Dockers\footnote{https://www.docker.com/}, a container platform that encapsulates an application in a stand-alone package.   The evaluation uses three graph types (the Albert and Barabasi model~\cite{barabasi2016network},  and real-world topology from CAIDA~\cite{CAIDA} and from the BRITE tool~\cite{BRITE}) for representing the network. The authors analysed the edge-nodes and the most suitable placements in the evaluation. From our point of view, however, the type of application used and the relationships among containers is not clear. 

EdgeCloudSim~\cite{EdgeCloudSim} is simulator based on CloudSim~\cite{cloudsim}, which is one of the most referenced simulators  in the field of cloud computing. Sonmez et \emph{al}. introduced functionalities such as mobility models, network link models and edge server models  to represent more realistic scenarios. Thus, new additional results were provided such as the LAN delay, number of failed tasks due to mobility and average number of mobile clients in a specific location. They presented a scenario with three configurations: one tier, two tiers, and two tiers with an edge orchestrator. The edge orchestrator entity controls the selection of the tier in each possible task execution. This simulator incorporates new functionalities relative to the original but is restricted in the taxonomy definition and how the mobility is defined. The type of results is also limited to the CloudSim version. 

iFogSim~\cite{ifogsim} is a CloudSim extension that supports the management of edge-network entities and the evaluation of allocation policies. The infrastructure is defined by a set of entities:\emph{ fog devices }(or fog nodes), \emph{sensors}, \emph{tuples} (such as a network link) and \emph{actuators}. The application is modelled as a directed graph with \emph{modules} (representing computational resources), \emph{edges} (a data dependency between application modules), and \emph{loops} (defining a sequence of \emph{edges} that should be monitored along the simulation to compute the response time. 
 In the article, the authors present two placement strategies that we describe in detail in the evaluation section: \emph{cloud-only placement} and \emph{edge-ward placement}. They introduce the simulator with two case studies: a latency-sensitive online game (namely, the EGG Tractor Beam game) and intelligent surveillance through distributed camera networks. Based on the iFogSim simulator, we use the application model in our simulator, introducing new improvements in the API, and we compare our results using the first case study and the two placement strategies as explained in the article.

\section{YAFS architecture}

YAFS uses a generic library for the generation of discrete event simulation scenarios called Simpy\footnote{\url{https://simpy.readthedocs.io}}. The Simpy library contains functions for the definition of processes (active components) and shared resources (such as network links and queues). It performs the execution of the simulation in three modes: as fast as possible, in real time, or manually stepping through events. 
It can also halt the simulation in case of lack of interaction or with a fixed step size.  Simpy is a robust and stable DES implementation that we use to implement  functions to control the atomic processes behind a fog domain: the transmission of workloads among network links, the computation of processes in fog nodes, and others issues that we describe below. 

\begin{figure}[!h]
	\centering
	\includegraphics[width=.8\linewidth]{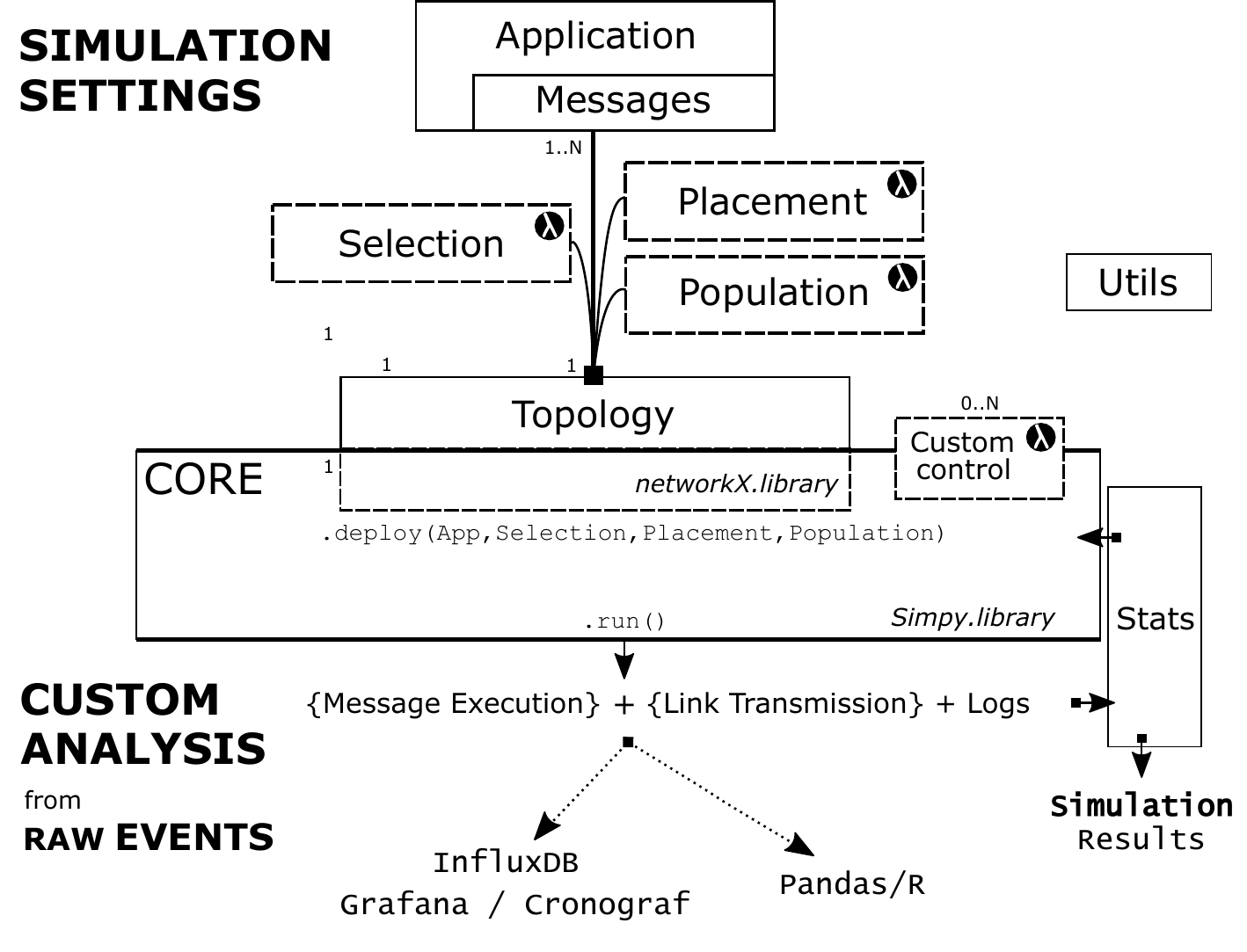} 
	\caption{The YAFS architecture is defined by six main classes: Topology, Core, Application, Selection, Placement, and Population.}
	\label{fig3}
\end{figure}

YAFS is defined by six main classes: core, topology, selection, placement, population, and application. Figure~\ref{fig3} shows the relationships among them. 
\emph{Core} class integrates the rest of the fog scenario definitions and manages the simulation execution controlling  the cycle of life of processes, including the customized policies: selection, placement, population and custom controls. 
The main element of the \emph{Core} class is the \emph{Topology}. The topology structure is accessed by the rest of the classes through the control of the \emph{Core} class. 
Simulation processes such as  \emph{Selection}, \emph{Placement}, \emph{Population}  are integrated into the \emph{Core} class to provide orchestration and selection of processes, allocation of software modules in the entities of the structure; and allocation and characterization of workloads, respectively.  As shown in Fig.~\ref{fig3}, the classes with a white lambda on a black circle symbol can interact dynamically  along the simulation execution.  \emph{Core} class gathers  events (such as message transmission and message execution) and stores them in a raw format. \emph{Stats} and \emph{Metrics} classes implement several functions to compute common measures such as the average response time, link latency, and resource utilisation.

In the following sections, we examine the topology and entity modelling, the application model, the internal structure of the DES processes and the generation of the results. API documentation and a tutorial are available on-line\footnote{\url{https://yafs.readthedocs.io/en/latest/}}, providing further detail.

\subsection{Topology and Entity modelling} 

We represent the relationships among network entities of a fog computing scenario  through a graph model where the nodes are network elements, routers, endpoints, fog nodes or similar elements and the vertices are the network links. As mentioned, we can apply complex network theory to this model.  We use the NetworkX~\cite{nx} library with functionalities for manipulation, visualization and extensively tested implementations. This library can import CAIDA~\cite{CAIDA} and BRITE~\cite{BRITE} topologies, and it supports graph formats such as JSON, GML, GEXF, Pickle, GraphML, and Pajek.

The mandatory attributes to define a fog node are an identifier (ID), the number of instructions performed per unit of time (IPT) and the memory capacity (RAM).  Developers can include other, customized tags to define the topology entities. In Fig.~\ref{declarationFogNodes}, we include two JSON-based definitions of nodes: one with a range of power consumption and a coordinate value and one that contains only the mandatory attributes. YAFS supports a flexible definition of entities in the same scenario. 
Using customized attributes, we can represent logical relationships such as virtualization, containers, microservices, and serverless functions using nodes and vertices.

\begin{figure}[!h]
	\centering
	\begin{lstlisting}
	{
	"id": 120, "RAM": 1, "IPT": 530,
	"POWERmin": 574,
	"POWERmax": 646
	"coordinate": {"lat":39.30, "long":3.34}
	},
	{
	"id": 12, "RAM": 10, "IPT": 100
	},
	\end{lstlisting}
	\caption{Definition of two fog nodes using a JSON-based representation.}
	\label{declarationFogNodes}
\end{figure}%


The definition of link attributes is similar. A network link has two mandatory attributes: bandwidth (BW) and link propagation (PR).

Finally, a simulation contains a unique \emph{topology} class. As mentioned, this class  is a graph-based representation where a determined number of applications and their corresponding policies are deployed.  That is, each application has a unique policy of allocation of resources (\emph{placement}), a policy of selection and orchestration of services (\emph{selection}), and a variation of the workload (\emph{population}). Furthermore, we can deploy \emph{customized controls} that  dynamically interact with the application and the simulation variables (such as failure generation or to improve the computational capacity of a node).

\subsection{Application model}

The application model is the same as that of iFogSim~\cite{ifogsim} and is based on a distributed data flow (DDF)~\cite{ddf}. An application is defined by modules that run services and messages (or dependencies) among modules. Thus, a DDF is  represented with a  directed acyclic graph where nodes are modules that perform one action on the incoming data and edges denote interoperability between modules. This application representation enables the partitioning and scaling of an application, which is useful for real program models such as microservices~\cite{Dragoni2017} and serverless~\cite{Baldini2017} paradigms.

We adapted the application definition with regard to the iFogSim approach to complying with our design principles: independence of the results and ease of export and reuse functions. This fact is reflected in each definition phase of an application: modules, dependencies, messages and results. The mandatory attributes of a message are \textit{instructions} and \textit{bytes}.  The instructions affect the service time, and the bytes affect the transmission time.   

In YAFS, all types of modules are defined with the same methods.  iFogSim authors use the term  \emph{dependency} to represent the relationship between modules, and these modules do not start the execution until they receive a message; instead, we use the term  \emph{message}. These messages can be used for other applications that have the same modules. The transfer of messages indicates  how to transform a type of input message into another output message. In YAFS, all transfers are defined, including the generation of messages in sensors or the reception  in actuators and the generation within modules (periodic messages). The decision to transmit a message within a module is also implemented, with two methods: fractional selectivity and broadcasting. The latter allows message transmission to all replicated modules.  Finally, in YAFS, the response time is obtained independently of the declaration of \emph{loops} (an internal control of iFogSim for monitoring tasks of sequence of dependencies between modules of an application), i.e., in iFogSim, if a loop is not declared before the simulation, the execution times of a sequence of dependencies between modules cannot be measured. 

To understand the differences between iFogSim and YAFS, we implement the application used in the first case study of iFogSim: the EGG Tractor Game (Fig.~\ref{applicationEGGapplication}). The game consists of three modules: \emph{client}, \emph{concentration calculator} and \emph{coordinator}, performing processing of the messages generated in the \emph{EGG sensor}; some results are visualized in the \emph{Display} actuator.  The modules are defined in lines 9-11; modules that will be workload sources or simple sensors are defined as \texttt{sources} and \texttt{sinks}. They are necessary only to define the application. The messages are defined in lines 11-13. The following attributes are required: name, module source, module destination, instructions, and bytes. Finally, the remaining lines define the transmissions. This is where we define how a message is transformed into another and how a message is sent between modules (through a distribution, a selection or a broadcast process).
The placements of workloads (source entities) are defined in the population policy.

\begin{figure}[!h]
	\centering
	\begin{lstlisting}
	a = Application(name="EGG_GAME")
	a.set_modules([
	  {"EGG":{"Type":Application.TYPE_SOURCE}},
	  {"Display": {"Type": Application.TYPE_SINK}},
	  {"Client": {"RAM": 10, "Type": Application.TYPE_MODULE}},
	  {"Calculator": {"RAM": 10, "Type": Application.TYPE_MODULE}},
	  {"Coordinator": {"RAM": 10, "Type": Application.TYPE_MODULE}}
	])
	m_egg = Message("M.EGG", "EGG", "Client", instructions=2000*10^6, bytes=500)
	m_sensor = Message("M.Sensor", "Client", "Calculator", instructions=3500*10^6, bytes=500)
	m_player_game_state = Message("M.Player_Game_State", "Calculator", "Coordinator", instructions=1000*10^6, bytes=1000)
	# ...
	a.add_source_messages(m_egg)
	dDistribution = deterministicDistribution(name="Deterministic", time=100)
	a.add_service_source("Calculator", dDistribution, m_player_game_state) 
	a.add_service_source("Coordinator", dDistribution, m_global_game_state)
	a.add_service_module("Client", m_egg, m_sensor, fractional_selectivity, threshold=0.9)
	a.add_service_module("Client", m_concentration, m_self_state_update, fractional_selectivity, threshold=1.0)
	# ...
	\end{lstlisting}
	\caption{Definition of EGG Game application presented in iFogSim~\cite{ifogsim} using our API.}
	\label{applicationEGGapplication}
\end{figure}%

%

\begin{figure}[!h]
	\centering
	\includegraphics[width=.84\linewidth]{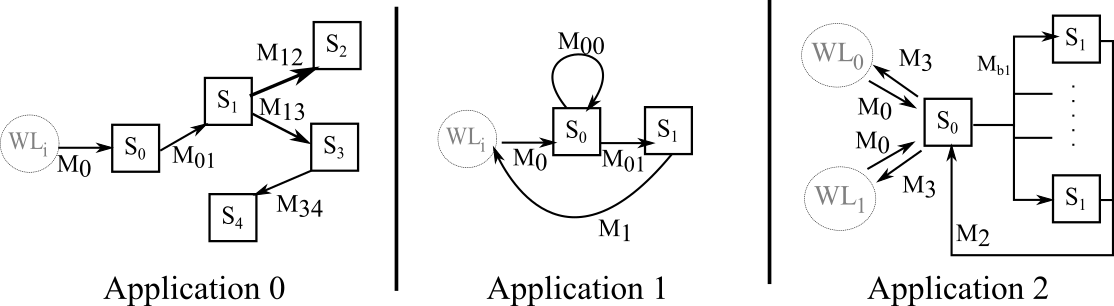} 
	\caption{Three application types with a simple message passing, a loop message and a broadcast message.}
	\label{fig3apps}
\end{figure}

Our implementation includes additional types of applications that can be modelled and their interactions with the workload generators or users with regard to other simulators. Figure~\ref{fig3apps} shows three application types.  Each application is set up by software modules represented with boxes. \emph{Application 0} presents a hierarchical structure where the messages (identified with $M{ij}$) trigger other messages. In the example,  post-execution of $M_{01}$ triggers $M_{12}$ and $M_{13}$. In  \emph{application 1}, we can observe a self-message and an interaction with the user from other module. In the last case, \emph{application 2}, there is a broadcasting message ($M_{b1}$) that reaches all $S_1$-module deployments.  Each $S_1$-module returns a message that is addressed by $S_0$-module and finally returns a response to a specific user.  This last feature enables the return of responses to the initial claimant. 

The supplementary material includes several examples of applications. In one of them, we define an application based on the structure of  \emph{application 0} (Fig.~\ref{fig3apps}) and the correspondent scenario (population policy, allocation policy and infrastructure) using a specific JSON-based syntax. The use of JSON files provides a common framework for the definition of Fog scenarios where non-expert developers can easily design experiments.


\subsection{Dynamic policies}
The \emph{Selection}, \emph{Placement} and \emph{Population} classes dynamically generate the events in the scenario. The first class chooses the entity that performs the execution of application modules; hence, it routes the workload.  The \emph{Placement} class determines the allocation of each application module. The \emph{Population} class allocates the workload generators in the network entities. These classes possess two main interfaces: an initialisation function and a function invoked according to a customized  temporal distribution. The initialisation function prepares the allocation of modules and workloads on topology entities. 

To illustrate these type of processes, we describe the population definition where we map workload generators in the entities of the infrastructure. This procedure requires three steps. First, we need to choose the type of message that is generated from the workload sources. The messages are defined in the \emph{Application} class, and they are requested to perform the execution of application modules.  Second, we have to define the temporal distribution. Finally, we have to associate how many of these generators we wish to have in the nodes. 
We have included an example of population criteria in Fig.~\ref{populationAlgorithm}). Lines 1 and 2 define two temporal distributions. The first one starts at 3000 time units, and from that point in time, it triggers an activation every 300 time units. The second one triggers an activation every 10 time units.  Line 4 generates an instance of a predefined extended class of \emph{Population}.  In \emph{application A}, there are two types of modules: workload sources and workload sinks (similar to sensors and actuators). Thus, Lines 5 and 6, through JSON-based syntax, define the allocation of each sink module (in this case, we incorporate all sink modules, and we duplicate the number in the same entity), and the allocation of each workload source with a distribution and a type of message. 

We can extend the \emph{Population} class to model more complex scenarios (Lines 8-11). In this sense, the new instance (Line 9) of an extended \emph{Population} operation follows the distribution defined in Line 1. This DES process starts generating workload sources at a certain time point, after which it is activated every 300 time units.  In each activation, it generates a new workload source with the characteristics defined in Line 11, and it is assigned in each entity defined in the array (\emph{top20Devices}). Sink modules are generated in the initialisation phase. A simplified version of this \emph{Evolution} class is shown in Fig.~\ref{especificacionClasePopulation}). In these types of processes, there is a mandatory function called \texttt{initial\_allocation}  and, optionally, a function called  \texttt{run} that is invoked dynamically according to the distribution. Internally, functions defined in the \emph{Core} class are used for our modelling, such as \emph{deploy\_source} or \emph{deploy\_sink}. Note that the \emph{sim} variable is the instance of the simulation. We can control  the topology and the rest of the DES processes, together with the simulation execution. This variable is created in the \emph{Core} class.

The rest of the classes (selection, placement and customized processes) present a similar structure and behaviour. We omit additional examples to avoid redundancy.

%

\begin{figure}[!h]
	\centering
	\begin{lstlisting}
	delayActivation      = deterministicDistributionStartPoint(3000,300,name="Deterministic")
	periodicActivation = deterministicDistribution(name="Deterministic", time=100)
	
	popA = Statical(name="StaticalPop")
	popA.set_sink_control({"id": a_id_fog_device, "number": 2, "module": appA.get_sink_modules()})
	popA.set_src_control({"number": 1, "message": appA.get_message("M.Action"), "distribution": periodicActivation})
	
	top20Devices = [''array_ids_fog_devices'']
	popB = Evolution(top20Devices, name="DynamicPop", activation_dist=delayActivation)
	popB.set_sink_control({"model": "actuator-device","number":2,"module":appB.get_sink_modules()})
	popB.set_src_control({"number": 1, "message": appB.get_message("M.Action"), "distribution": periodicActivation})
	\end{lstlisting}
	\caption{Declaration of two population policies: one static (\emph{popA}) and the other dynamic (\emph{popB}).}
	\label{populationAlgorithm}
\end{figure}%

\begin{figure}[!h]
	\centering
	\begin{lstlisting}
	class Evolution(Population):
	  def __init__(self,listIDEntities,**kwargs):
 	    #initialisation of internal variables...
	    super(Evolution, self).__init__(**kwargs)
	
 	  def initial_allocation(self, sim, app_name):
 	    #dealing assignments...
 	    sim.deploy_sink(app_name, node=fog_device, module=module)
	
	  def run(self, sim):
 	    #dealing assignments: msg, distribution and app_name.
 	    id = ... # listIDEntities.next
 	    idsrc = sim.deploy_source(app_name, id_node=id, msg=..., distribution=...)
	\end{lstlisting}
	\caption{Structure of a population class with three mandatory functions: \emph{init}, \emph{initial\_allocation} and \emph{run}.}
	\label{especificacionClasePopulation}
\end{figure}%

\subsection{Results}
    
There are two types of events recorded (namely, task executions and network transmissions), but users can record specific metrics with customized DES processes.
The results are stored in two CSV files.

When a node performs the work associated with a message, the simulator records the following attributes: \textit{id, type, app, module, message, DES.src, DES.dst, TOPO.src, TOPO.dst, module.src, service, time\_in, time\_out, time\_emit, time\_reception}. Specifically, \textbf{ \textit{id}} is an incremental integer value that remains constant during message propagation or transformation in other messages of the application. This approach allows controlling when an application partitioned in modules ends the execution of its complete service. The attribute  \textbf{\textit{type}} identifies the type of module (computational or sink). The attribute  \textbf{\textit{app}} identifies the application (the name attribute). \textbf{\textit{module}} identifies the application module (the name attribute) that performs the service. \textbf{\textit{message}} identifies the message (the name attribute). \textbf{\textit{DES.src}} and \textbf{\textit{DES.dst}} are the identifiers of the DES processes that send and receive the message, respectively.  \textbf{\textit{TOPO.src}} and \textbf{\textit{TOPO.dst}} are the identifiers of the topology entities where the modules are deployed.  \textbf{\textit{module.src}} identifies the application module that sends the message. 
The \textbf{\textit{service}} attribute can have a None value (if the message record comes from a workload source) or a numerical value corresponding to the service time.

\begin{figure}[!h]
	\centering
	\includegraphics[width=.8\linewidth]{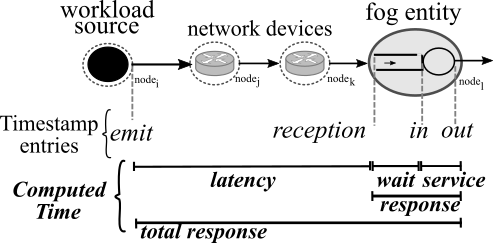} 
	\caption{Logged and computed times present in the fog node computation of a message.}
	\label{fig3b}
\end{figure}

Figure~\ref{fig3b} shows the four timestamps involved in the transmission of a message from the source  to the destination entity where the software module performs the action. 
The label \textbf{\textit{time\_emit}} is the value that represents the emission time of a message in a module source. The label \textbf{\textit{time\_reception}} represents the recorded  time when a message arrives to the destination module. When the message arrives, it is enqueued; finally, we record the entry and the exit of the service (\textbf{\textit{time\_emit}} and \textbf{\textit{time\_out}}, respectively).
The service time is the division of \textit{instructions} (message attribute) between \textit{instructions per time} (entity attribute).
These times are used to compute  useful measures such as the latency, waiting time,  response time and total response time (see computed times in Fig.~\ref{fig3b}). We show a sample of those records in Fig.~\ref{csv1}. Using the timestamps entries of the first requests (Fig.~\ref{csv1}), $time_{in}$ is 104.005, $time_{out}$ is 105.9994, $time_{emit}$ is 100.0, and $time_{reception}$ is 104.0005; we compute the latency as 104.0005-100.0, the waiting time as 104.005-104.005, the service time as 105.9994-104.005, the response as 105.9994-104.005, and the total response as 105.9994-100.00.

\begin{figure}[!h]
	\centering
	\scriptsize
	\begin{lstlisting}
	id,type,app,module,message,DES.src,DES.dst,TOPO.src,TOPO.dst,module.src,service\\
	,time_in,time_out,time_emit,time_reception
	1,COMP_M,EGG_GAME,Client,M.EGG,0,258,4,3,EGG,1.9994,104.005,105.9994,100.0,104.0005
	2,COMP_M,EGG_GAME,Client,M.EGG,2,259,7,6,EGG,1.9994,104.005,105.9994,100.0,104.0005
	3,COMP_M,EGG_GAME,Client,M.EGG,4,260,10,9,EGG,1.9994,104.005,105.9994,100.0,104.0005
	...
	\end{lstlisting}
\caption{Fog node computations recorded in a CSV file for subsequent analysis of the results.}
\label{csv1}
\end{figure}%

In a network transmission, YAFS records the following attributes: \textit{id, type, src, dst, app, latency, message, ctime, size, buffer}.  Here, \textbf{ \textit{id, type, app,}} and  \textbf{\textit{message}} take the same values as mentioned previously.   \textbf{\textit{src}} and \textbf{\textit{dst}} are the identifiers of the topology entities that send and receive the messages.  \textbf{\textit{ctime}} is the simulation time when the action is performed. \textbf{\textit{size}} is the size of the message. Finally, \textbf{\textit{buffer}} is an integer value that represents the number of messages in the whole network that are waiting for a link service. Consequently, a link can send only  one message at a time, and messages have to wait for unused slots. This value is an indicator of network saturation and is updated in each record.  In the case shown in Fig.~\ref{fig3b}, the transmission of the message from the workload source to the fog node generates three network transmission records. We show a sample of these types of records in Fig.~\ref{csv2}. 


\begin{figure}[!h]
	\centering
	\scriptsize
	\begin{lstlisting}
	id,type,src,dst,app,latency,message,ctime,size,buffer
	1,LINK,4,3,EGG_GAME,4.000005,M.EGG,100,500,0
	2,LINK,7,6,EGG_GAME,4.000005,M.EGG,100,500,1
	3,LINK,10,9,EGG_GAME,4.000005,M.EGG,100,500,2
	...
	\end{lstlisting}
	\caption{Network link transmissions are recorded in another CSV file.}
	\label{csv2}
\end{figure}%

We have implemented some common methods (in the \emph{Stats} class) to obtain more complex measures using the Pandas library~\cite{pandas}. Pandas is an open source library with several data analysis tools. To illustrate this data analysis, we include the next example in Fig.~\ref{resultados}.  The first approach follows the idea of \textit{sequences} defined in iFogSim. From a sequence of messages (line 2), the \emph{showResults} function provides the same results as iFogSim (line 4). In addition, we can perform more complex analysis. For example, we can compute the average latency each 300 units of time. To obtain these values, we use the Pandas time series functionalities to sample the records in that time period and to apply the average function on latency values (lines 6-8), where \texttt{df} (a \textit{dataframe}) contains the CSV data. 

%
\begin{figure}[!h]
	\centering
	\begin{lstlisting}
	simulation_time = 100000
	time_loops = [["M.EGG", "M.Sensor", "M.Concentration"]]
	s = Stats(defaultPath=path+"Results")
	s.showResults(simulation_time, time_loops=time_loops)
	
	s.df["date"]=s.df.time_in.astype('datetime64[s]')
	s.df.index = s.df.date
	print s.df.resample('300s').agg(dict(time_latency='mean'))
	\end{lstlisting}
	\caption{Analysis of the simulation results.}
	\label{resultados}
\end{figure}%

\section{Evaluation}


In the first section, we compare YAFS and iFogSim simulators  in terms of performance and results using an application case defined in iFogSim~\cite{ifogsim}.  In the second section, we analyse the convergence of both simulators using the same experiments.

It is important to note that the results are not equal between both simulators, although we try to use similar settings. The definition of attributes is different in both simulators. These cases are the following: I) iFogSim uses the measure of MIPS in its computational devices. YAFS uses IPT  (\textit{instructions per time}. II)  In iFogSim, the attributes of a link are included in the fog node using terms such as \textit{upBW} and \textit{downBW}, and there is another latency value in the connection between modules (i.e., \textit{eegSensor.setLatency(6.0)}). In our case, the BW is defined in the link and has the same value in both directions. In addition, we define the propagation time, which is not included in iFogSim III)  In iFogSim, a message has attributes such as \textit{tupleCPULength} and \textit{tupleNwLength}, corresponding to the number of millions of instructions and bytes, respectively.  In any case, temporal distributions are the same in the experiments, and we try to use similar values in the previously described attributes.

\subsection{Comparison with iFogSim}
\begin{figure}[!h]
    \includegraphics[width=.4\linewidth]{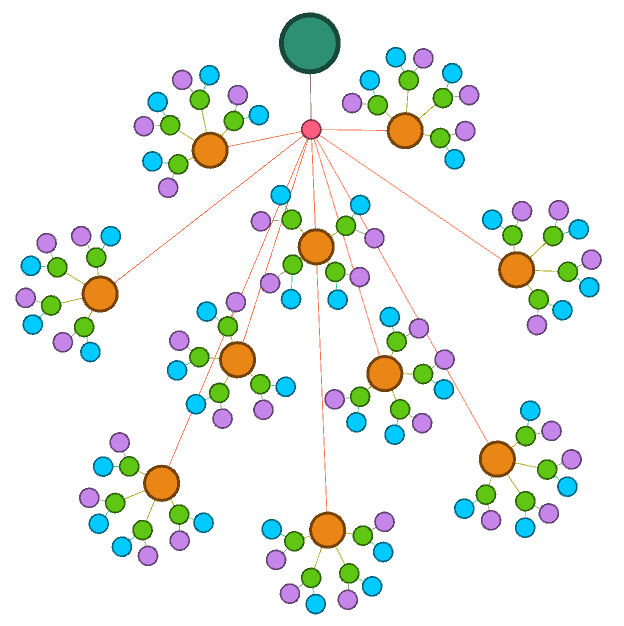}
	\hspace{1.0 cm}
	\includegraphics[width=.45\linewidth]{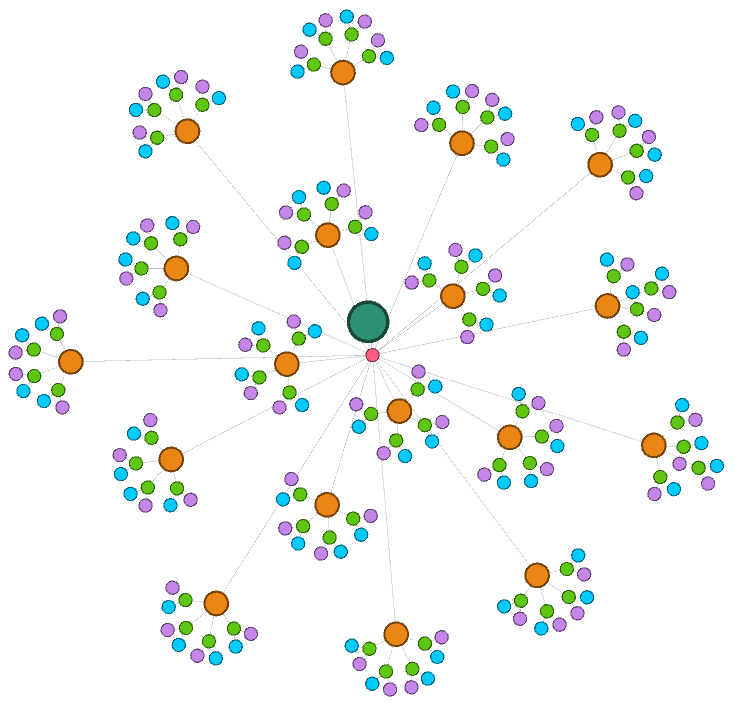}
	\caption{Network topologies with 10 fog devices (left figure) and 18 fog nodes (right figure). The large green node represents the cloud entity, pink nodes are the proxies, orange nodes are the gateways or fog devices, and small green nodes are client devices with one sensor (purple) and one actuator (blue).}
	\label{fig5}
\end{figure}

We use the first case study presented in the iFogSim paper (namely, the EGG Tractor Beam game) for the comparison between both simulators. This application consists of 3 modules:  \emph{client}, \emph{concentration}, and \emph{coordinator}, and the experiment deploys the modules in a hierarchical three-based topology with a cloud entity that is linked to a gateway where all fog devices are connected. The network can be scaled from the gateway device generating several subgroups. Fig.~\ref{fig5} represents an example of two topologies with 10 and 18 gateway subgroups. 

We analyse two different placement strategies: a \emph{cloud-only placement} (cloud policy) where all modules are deployed in the cloud entity and an \emph{edgeward placement} (edge policy) where the modules are deployed in fog devices (orange nodes in Fig.~\ref{fig5}). Both strategies are explained in the iFogSim paper. From the simulation, we analyse the following data:  execution time and response time. In addition, we vary the number of fog nodes: 4, 8, 12 and 16. The simulation is executed in a machine with 8 i7-cores running at 3.745 GHz  with 8 GB RAM. Because of the stable convergence of both simulators, as described in Section 4.2, we performed each experiment only once.

\begin{figure}[!h]
	\centering
	\includegraphics[width=.85\linewidth]{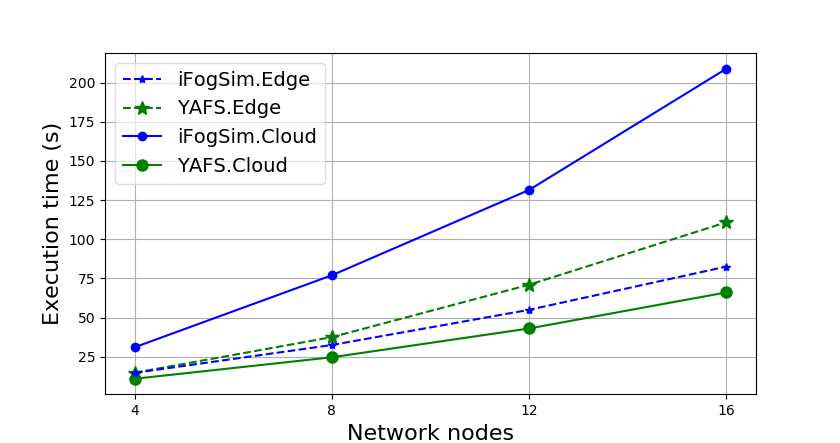}
	\caption{Execution time comparison between \emph{cloud} and \emph{edge} policies with a different number of gateways: 4, 8, 12 and 16.}
	\label{fig6}
\end{figure}

Figure~\ref{fig6} shows the execution time in both policies with regard to the increment of fog nodes. Blue lines are the results of iFogSim and green lines, YAFS. Circle marks correspond to \emph{cloud policy} and star marks to \emph{edge policy}. At first glance, the behaviour of both simulators is quite similar, but we can appreciate some differences: I) In the \emph{cloud policy}, a greater number of transmissions must be made since all messages go through more network links to compose the cloud entity. This volume of traffic generates a saturation in the network that affects the iFogSim runtime; II) \emph{edge policy} generates more application modules; there are more DES processes to control each module, and this fact slightly affects the YAFS runtime.  An increment of the simulation runtime is reasonable as more modules are controlled; however, the saturation of the simulated system not should affect the simulator itself.

\begin{figure}[!h]
	\centering
	\includegraphics[width=.8\linewidth]{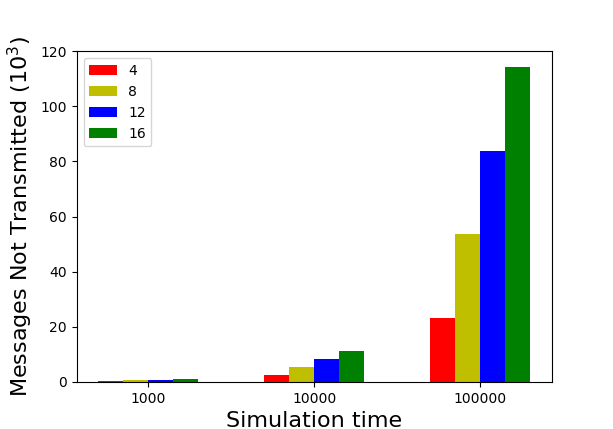}
	\caption{Number of messages enqueue (waiting) by network saturation  using YAFS with \emph{cloud policy} in increments of fog devices (4, 8, 12 and 16).}
	\label{fig7}
\end{figure}

The network is saturated with the parametrisation of the \emph{cloud policy} experiment. The saturation is greater when there are more network devices and is proportional between different gateway subgroups. In Fig.~\ref{fig7}, we represent the total number of  messages waiting for the service in each level of fog nodes using YAFS. iFogSim does not provide this measure. 

Another comparison is the latency time of the application. In this experiment, the latency is calculated as the sum of transmissions among the three modules, and  the response times is included:  the EGG sensor $\leftrightarrow$ client module,  client $\leftrightarrow$ coordinator module and coordinator  $\leftrightarrow$ client module. 
A substantial difference between the simulators is the need to indicate the sequence of messages in iFogSim before simulation execution. In YAFS, this step is not necessary; those latencies are calculated post-simulation.

With the  \emph{edge policy}, the clients and fog devices entities always have the same network distance, and the response time has to be constant and independent of  the number of fog nodes. Table.~\ref{tablaLatency} shows that the convergence of the latency time is better in YAFS. Note that the seed of the random numbers is always the same in each increment of gateways in both simulators. This seed changes only with the number of messages and the simulation time.

With the \emph{cloud policy}, the coordinators are allocated in the cloud entity. Subsequently, more messages are transmitted across the network and must pass through the same link. In both cases, the latency presents an exponential trend (Table.~\ref{tablaLatency}).  The parametrisation avoids the network saturation in the iFogSim execution with 4 fog nodes.

\begin{table}[]
	\small
	\centering
	\begin{tabular}{cc|cc|cc}
		& Simulator   &       \multicolumn{2}{c}{YAFS}     &    \multicolumn{2}{c}{iFogSim}       \\\hline
		& {\small Simulation Time } & $10^3$ & $10^5$  & $10^3$ & $10^5$ \\\hline\hline
		\multicolumn{6}{c}{Cloud Policy} \\\hline		
		& 4  &   725.7662        &    59778.5152    &  225.5632     &   225.5609      \\
		& 8  &   847.6964          &    78614.0355    &  440.5626    &   28074.4655       \\
		& 12 &   1100.2678        &    85519.9283    &  536.8919     &  40334.4660    \\
		\multirow{-4}{*}{\rotatebox[origin=c]{90}{fog nodes}}	& 16 &   459.7284         &    89119.8638    &  569.4413     & 44597.4717\\\hline\hline
		\multicolumn{6}{c}{Edge Policy} \\\hline		
		& 4  &   11.2744            &  11.2696       &    30.5536    & 31.6587         \\
		& 8 & 11.2691                &   11.2667      &  31.4244     &  31.6954        \\
		& 12  &  11.2671             &   11.2654  &  30.8828      &  31.6785       \\
		\multirow{-4}{*}{\rotatebox[origin=c]{90}{fog nodes}}& 16 & 11.2661 & 11.2649 &  30.2859  & 31.6964\\ 
	\end{tabular}
	\caption{Latency time with two different policies: \emph{cloud} and \emph{edge} varying the number of fog nodes and the simulation time}
	\label{tablaLatency}
\end{table}

\subsection{Convergence}
We analyse the convergence of YAFS using the same example of iFogSim as in Section 4.1.  In this case, we use the \emph{edge policy} since it is a stable configuration of the system.
We run both simulators 50 times with a simulation time of 10,000 units in each fog node configuration (4, 8, 12 and 16) to compute the latency time.

In this experiment, two factors change the precision of the latency time: the simulation time and the number of fog nodes. The first factor exhibits coherent behaviour in simulation experiments but in this case is constant. The second factor, the number of fog nodes, affects the number of transmitted messages; then, it statistically increases the number of samples. We can expect a reduction in the variance in each fog node increment.  In Table~\ref{convergencia}, we include the numerical values (mean, variance, minimum, and  maximum) of each simulator.  We observe the reduction of the variance in each experiment. The divergence between the different ranges may be due to differences in the configuration of the experiment in each simulator, but YAFS is slightly more stable.

\begin{table}[]

	\centering
	\scalebox{0.85}{
	\begin{tabular}{cc|cccc||cccc }
		&  & \multicolumn{4}{c||}{YAFS} & \multicolumn{4}{c}{iFogSim} \\\hline
		&  &   mean  & var.    & min.    & max.   &  mean   & var.    & min.    & max.   \\\hline
		& 4  &    11.2697  &  4.646e-07    & 11.2683    & 11.2706        &  31.6771   &  0.0414   & 31.2096    & 32.0984   \\
		& 8  &    11.2664  &   9.704e-07   & 11.2655    &  11.2668       &  31.6699  &  0.0203   & 31.3082    &  31.9101    \\
		& 12 &  11.2654   &  4.018e-08   & 11.2649     &  11.2656      &  31.6564    &   0.0156  &    31.3886 &   32.0163 \\
		\multirow{-4}{*}{\rotatebox[origin=c]{90}{fog nodes}} 
		&  16&  11.2648   &  2.064e-08   & 11.2642    & 11.2650   		   &  31.6598   &  0.0169   &  31.3923   &   31.9032
	\end{tabular}
	}
	\caption{Convergence of YAFS and iFogSim using 50 samples with an \emph{edge policy} configuration and different number of fog nodes with 10,000 time units of simulation.}
	\label{convergencia}
\end{table}

\section{Three  complex scenarios}

In a second experimental phase, we highlight selected YAFS features, and we implement three dynamic IoT scenarios: allocation of new modules, failures on the infrastructure, and user mobility.  

We analyse the behaviour of the results of these three experiments, but we have not been able to compare them with real cases.  To the best of our knowledge, there are no public data from these types of scenarios, and the real characterization of some of these studies is outside the scope of this article. Thus, we have used arbitrary values  (average values, distributions, etc.) to describe the expected behaviour of the results.

The first step is the definition of the network infrastructure (or topology). To illustrate the use of complex networks, we use the Graph Stream Generator library~\cite{GraphStream} to create a Euclidean random graph~\cite{Erdos}. This topology is the same for all three experiments (with a size of 400 nodes and 2242 edges) where the links have the same propagation speed (1 time unit) and  fog nodes can serve an unlimited number of modules.  We choose this type of graph since such graphs represent social relationships among individuals and have a high connection degree.  

The application consists of two modules: \emph{senders} and \emph{receivers}, and it has only one type of message. In this way, complex data analysis is avoided in the experiment. Initially, we randomly allocated 100 \emph{senders} in the topology, and the number of \emph{receivers} depends on each case study.  Each \emph{sender} generates a message  each 10 time units, and the service time of the \emph{receiver} is 0.0. Thus, the response time is equal to the latency time. To ensure accurate replication of the experiments, the seed of the random number generator is the same for all the experiments. 

In these three experiments, the results (latency times) are average values from the simple sequence between a \emph{sender} and a \emph{receiver}. The computation is similar to lines 6-8 from Fig.~\ref{resultados}, i.e., the value is the average aggregation of a set of values of a time period. The selection policy is based on the minimal path distance between a \emph{sender} and a \emph{receiver}.

We allocate the \emph{receiver} modules selecting the nodes with the biggest betweenness centrality of a graph~\cite{Newman}. A higher value of node centrality corresponds to a greater occurrence of the node as part of the shortest path between two other nodes.  The goodness-of-fit evaluation of this measure as an indicator to select a network device such as a fog node is not part of this study, but some analyses have been performed in previous studies~\cite{fmec2018}. All three experiments and results are available in the code repository in the example folder ~\footnote{https://github.com/acsicuib/YAFS/tree/master/src/examples}.

\subsection{Dynamic allocation of modules}
In the first scenario, we scale the number of \emph{receiver} modules. The objective is to observe how the latency time improves as this number grows. In the initialisation phase,  100 \emph{senders} are deployed with one \emph{receiver}, which is deployed on the node with  the highest betweenness centrality.  From time point 3000 of the simulation time, a new receiver is added with a period of 300 units. This process is repeated 19 times (a total number of 20 receivers are deployed). With this experimentation configuration, we observe that with 20 fog nodes, the response time tends to be stable. 

\begin{figure}[!h]
	
	\begin{subfigure}{1.\textwidth}
		\centering
		\includegraphics[width=.4\linewidth]{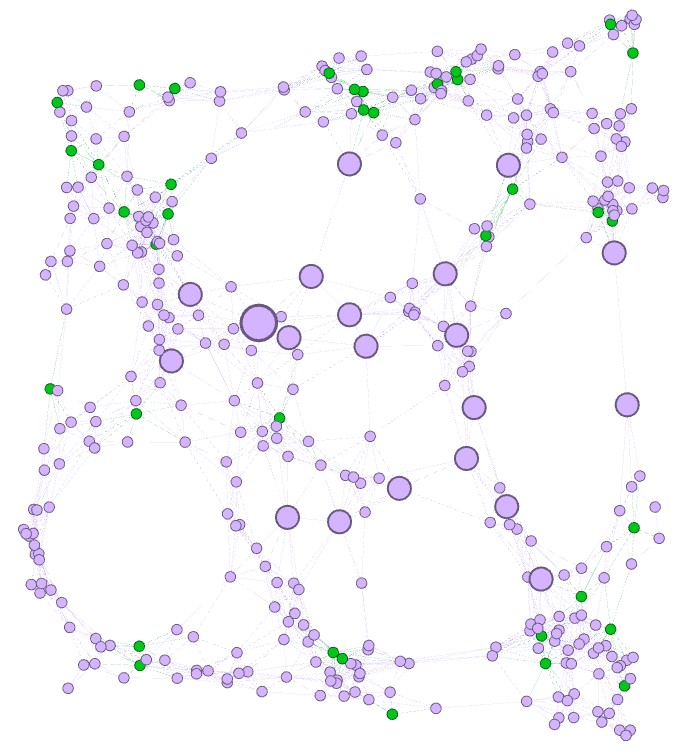}
		\caption{Euclidean random topology where the senders are allocated on green nodes and the size of the nodes represents the betweenness centrality.}
		\label{fig10a}
	\end{subfigure}
	\begin{subfigure}{1.\textwidth}	
		\centering
	    \includegraphics[width=.6\linewidth]{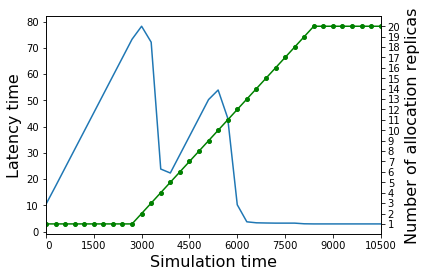}
		\caption{Evolution of the latency time (blue line) relative to the number of allocated fog nodes (green dotted line).}
		\label{fig10b}
	\end{subfigure}
     \caption{First scenario: dynamic allocation of modules.}
	 \label{fig10x}
\end{figure}

Figure~\ref{fig10a} represents the network, where green nodes contain the \emph{senders} and the size of the nodes represents the betweenness centrality.  The results of the execution are shown in Fig.~\ref{fig10b}. The blue line is the evaluation of latency time, and the green dotted line is the number of \emph{receivers} deployed.

We can observe that the network is saturated with only one receiver (from 0 to 3000 in simulation time) because the latency is continuously increased. From the first deployment (time point 3300), the latency is reduced due to a higher number of available receivers, and the messages are more evenly distributed across the network.  From the fifth module, however, the inclusion of new receivers along the network does not introduce any improvement since previous receivers still receive the workload.  Note that the selection process of receivers is based on the minimum path between a source and a destination node. Starting at the eleventh module, the allocations have an impact in the selection strategy, and the latency is again stabilised.

\subsection{Dynamic failures on network devices}
In the second scenario, we implement a dynamic failure of nodes where the failure rate is based on an exponential distribution.  The objective is to observe how the latency time worsens as this number of failures grows and consequently to show how the simulator can implement dynamic scenarios. In this experiment, we remove  only the fog nodes and other network entities. \emph{Sender} modules are not removed to ensure that the workload is the same throughout the simulation.

When a node fails, the node and its links are removed from the topology, which can affect the internal processes that the simulator handles. Thus, a new routing is computed for the messages that had a path through the failed node. If there is no other possible path, the simulator catches and records this outcome in a log.  When a removed node has waiting messages to be served, the messages are discarded, and the simulator records this case.

In the initialisation of this experiment, there are 100 \emph{senders} and 20 \emph{actuators} deployed. All of them are allocated in the same nodes as in the previous experiment.  The failures are generated from 500 time units and beyond with a mean of 100 time units. At the end of the experiment, the number of nodes available is 314, and the number of links is 1359, i.e., 86 nodes and 883 links are removed.

\begin{figure}[!h]
	\begin{subfigure}{1.\textwidth}
		\centering
		\includegraphics[width=.4\linewidth]{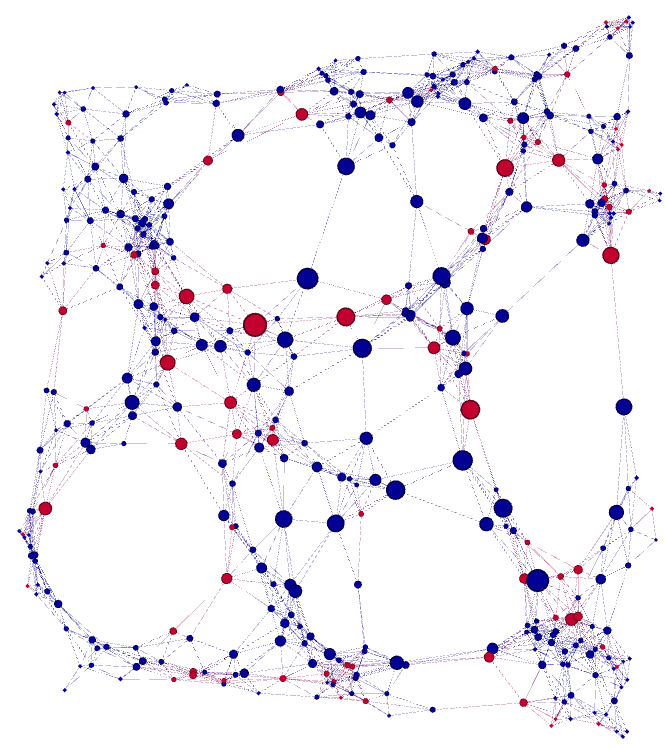}
		\caption{Network topology where red nodes are randomly chosen to be removed.}
		\label{fig11a}
	\end{subfigure}

	\begin{subfigure}{1.\textwidth}	
		\centering
		\includegraphics[width=.6\linewidth]{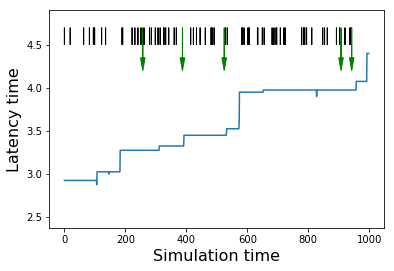}
		\caption{Evolution of the latency with failures on fog nodes (green arrows) and failures on network nodes (black lines).}
		\label{fig11b}
	\end{subfigure}

	\caption{Second scenario: dynamic failures on network devices.}
	\label{fig11}
\end{figure}

We represent the topology in Fig.~\ref{fig11a},  where red coloured nodes are randomly chosen to be removed during the simulation.  There are five red coloured nodes (fog nodes with allocated \emph{actuators}), which will be removed. 

In Fig.~\ref{fig11b}, we represent the evolution of aggregate latency times (samples are aggregated each 100 time units), and the failures are represented with black lines or green arrows in the upper part of the graph. A black line marks the failure of a network device, and a green arrow represents the failure of a fog device. As we can observe, the latency worsens as failures occur.

\subsection{Dynamic movement of message senders}

In the third scenario, we wish to characterise the movement of users (\emph{sender} modules) in the infrastructure. Therefore, \emph{receivers} are statically located in the nodes, and the allocation of \emph{senders}  is changed in periodic steps. As a result, for every period, we reduce the hop count by one between \emph{senders} and \emph{receivers}. The objective is to observe how the latency time improves and how the simulator can model scenarios with dynamic workloads.

In the initialisation phase, there are  100 \emph{senders} randomly allocated and  20 \emph{receivers}.  All the \emph{receivers}  are allocated in the node with the highest betweenness centrality. 
Every 400 time units, all the \emph{senders} are moved to the next nearest node with regard to the \emph{receiver} nodes. We use the shortest path function to compute the next node. As the links around the \emph{receiver} nodes receive many requests, we reduce the generation rate of the requests (100 time units). In addition, the selection policy of this experiment includes a round robin scheduler to select different \emph{receivers}.

\begin{figure}[!h]
		\begin{subfigure}{.45\textwidth}
		\centering
		\includegraphics[width=1.\linewidth]{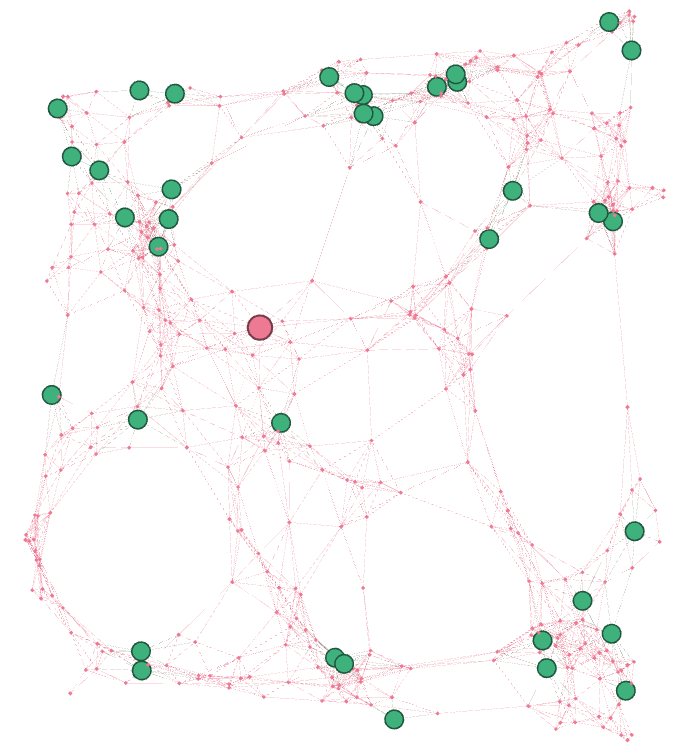}
		\caption{Initial localisation of \emph{sender} modules (green nodes) on the topology. Receivers are in the pink node.}
		\label{fig12a}
	\end{subfigure}
   \begin{subfigure}{.45\textwidth}
		\centering
		\includegraphics[width=1.\linewidth]{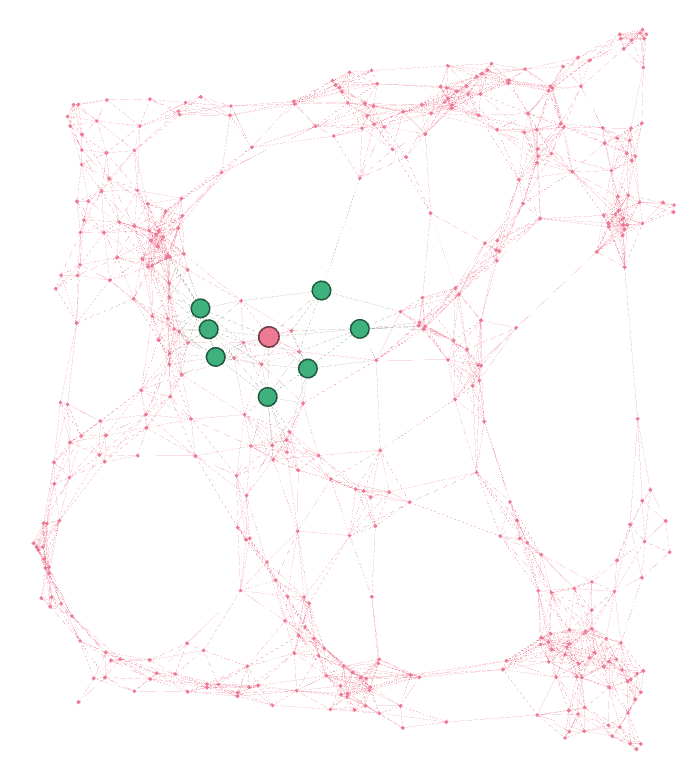}
		\caption{Final localisation of \emph{sender} modules.}
		\label{fig12b}
	\end{subfigure}
	\begin{subfigure}{1.\textwidth}
		\centering
	    \includegraphics[width=.7\linewidth]{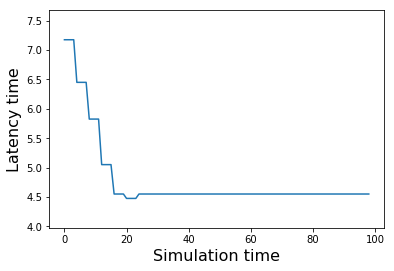}
		\caption{Evolution of the latency time during the simulation.}
	    \label{fig12c}
	\end{subfigure}
	\caption{Third scenario: dynamic movement of workload sources.}
	\label{fig12}
\end{figure}

Figure~\ref{fig12a} and Fig.~\ref{fig12b} represent the topology (same layout configuration than previous experiments). The green coloured nodes represent the initial localisation of the \emph{senders} (Fig.~\ref{fig12a}),  and at the end of simulation (Fig.~\ref{fig12b}). The pink coloured node contains the \emph{actuators}. 

The latency decreases at each step and ultimately converges at approximately 4.5 time units (Fig.~\ref{fig12c}).  The latency is obtained from the aggregation of the time series of events every 100 time units. Most of the \emph{senders} pass from the node with most closeness to the \emph{actuators} node in an average of 5 steps.  

In summary, the customisation of  temporal distributions and the structure of the YAFS engine enable a direct and flexible control of any type of event inside of the DES engine. Another notable aspect of the YAFS design is that it is based on a style of open programming, maximising the use of third-party libraries for delegating internal tasks such as the generation of topologies, visualization, or data analysis. For instance, as we show in experiments, we use complex network theory to perform several studies, and we export the topology to other graph formats for debugging and visualisation.

\section{Conclusion and future work}

We present a fog computing simulator for modelling novel and complex IoT domains. Our simulator, called YAFS, meets several design objectives: a light syntax, a user customised configuration of policies and a dynamic invocation of policies during the simulation, a definition of network topologies based on complex network theory, and the capacity to record computational and transmissions results  in a CSV format. This last point makes the simulator ideal for enhancing interoperability with third-party libraries, such as Grafana, for the creation of control panels to simulate monitoring infrastructures, or Panda or R for data analysis. In addition, the infrastructure and policies definitions can be done following a predefined JSON-based format, which simplifies use by non-expert programmers and facilitates the integration of  results of optimisation algorithms for the evaluation of fog placement proposals.

YAFS is a simulator based on discrete events where the management of shared resources and simulation control relies on the Simpy library, which was specifically designed to simulate  scenarios under Python.  The infrastructure is modelled using complex graphs, which allows efficient delegation of the infrastructure management including topological features,  that can be integrated into the customised policies. These policies are the definition of allocation modules in the fog nodes, the location of the workload generators (such as sensors or users), and  the selection of resources to compute tasks (including both path routing and orchestration and the scheduling of jobs). The application model is based on a distributed data flow (DDF) defined in the iFogSim simulator.  YAFS is available in a code repository\footnote{\url{https://github.com/acsicuib/YAFS}} containing the implementation of all previous cases in addition to several examples and a documented API.

Regarding the evaluation, we compare two policies (cloud and edge allocations) with iFogSim. In both policies, the convergence of the results is similar. However, the YAFS runtime is slightly better than that of iFogSim. As YAFS has more functionalities, we design three complex experiments that are not compared with iFogSim since they cannot be implemented under its API: in the first experiment, we create new fog nodes; in the second one, we dynamically simulate failures of devices; and in the third experiment, we represent the movement of workloads in the infrastructure. The results are consistent with the expected values in each experiment.

Future work will mainly cover the development of power-aware management policies, functions for controlling the computational capacity of the resources and improvements in the nomenclature.

\section*{Acknowledgements}
This research was supported by the Spanish Government (Agencia Estatal de Investigaci\'on) and the European Commission (Fondo Europeo de Desarrollo Regional) through grant number TIN2017-88547-P (MINECO/AEI/FEDER, UE).

\section*{References}

\bibliography{mybibfile}

\end{document}